\documentclass{blois}

\usepackage{amsmath}
\usepackage{enumitem}

\def\be{\begin{equation}}
\def\ee{\end{equation}}
\def\bea{\begin{eqnarray}}
\def\eea{\end{eqnarray}}

\newcommand{\Photo}{}

\newcommand{\gev}{\,\, \mathrm{GeV}}
\newcommand\MHp{M_{H^\pm}}
\newcommand{\br}{\text{BR}}

\begin{document}
\vspace*{4cm}
\title{The ``96 GeV excess'' in the N2HDM}

\author{T. Biek\"otter}
\addresss{DESY, Notkestrasse 85,
D-22607 Hamburg, Germany}

\author{M. Chakraborti}
\addresss{IFT (UAM/CSIC), Universidad Aut\'onoma de Madrid, 
Cantoblanco, E-28048, Spain}

\author{S. Heinemeyer}
\address{IFT (UAM/CSIC), Universidad Aut\'onoma de Madrid
Cantoblanco, E-28048, Spain\\
Campus of International Excellence UAM+CSIC, Cantoblanco, E-28049,
Madrid, Spain\\
Instituto de F\'isica de Cantabria (CSIC-UC), E-39005 Santander, Spain}

\maketitle\abstracts{
The CMS collaboration reported an intriguing $\sim 3 \, \sigma$ (local) excess
at $96\;$GeV in the light Higgs-boson search in the diphoton decay mode.
This mass coincides with a $\sim 2 \, \sigma$ (local) excess in the
$b\bar b$ final state at LEP. We present the interpretation of this
possible signal as the lightest Higgs boson in the 2 Higgs Doublet Model
with an additional real Higgs singlet (N2HDM). We show that the type II
and type IV (flipped) of the N2HDM can perfectly accommodate both
excesses simultaneously, while being in agreement with all
experimental and theoretical constraints. The excesses are most
easily accommodated in the type II N2HDM, which resembles the
Yukawa structure of supersymmetric models. We discuss the
experimental prospects for constraining our explanation
via charged Higgs-boson decays at the LHC or direct production
of the $\sim 96\,$GeV Higgs boson at a future lepton collider
like the ILC.
}

\section{Introduction}

The Higgs boson discovered in 2012 by ATLAS and
CMS~\cite{Aad:2012tfa,Chatrchyan:2012xdj} is
so far consistent with the existence of a
Standard-Model~(SM) Higgs boson~\cite{Khachatryan:2016vau}
with a mass of $\sim 125\,$GeV.
However, the experimental uncertainties on
the Higgs-boson couplings are (if measured already)
at the precision of $\sim 20\%$, so that there
is room for Beyond Standard-Model (BSM) interpretations.
Many theoretically well motivated extensions of the
SM contain additional Higgs bosons. In particular,
the presence of Higgs bosons lighter than $125\,$GeV
is still possible.

Searches for light Higgs bosons have been
performed at LEP, the Tevatron and the LHC.
Besides the SM-like Higgs boson at $125\,$GeV
no further detections of scalar particles
have been reported. However, two excesses
have been seen at LEP and the LHC
at roughly the same mass, hinting to a
common origin of both excesses via a new
particle state.
LEP observed a $2.3\, \sigma$ local excess
in the~$e^+e^-\to Z(H\to b\bar{b})$
searches\,\cite{Barate:2003sz}, consistent with a
scalar of mass ~$\sim 98\,$GeV, where the mass
resolution is rather imprecise due to the
hadronic final state. The signal strength
was extracted to be
$\mu_{\rm LEP}= 0.117 \pm 0.057 \; ,$
where the signal strength $\mu_{\rm LEP}$ is the
measured cross section normalized to the SM expectation
assuming a SM Higgs-boson mass at the same mass.

CMS searched for light Higgs bosons in the diphoton
final state. Run II\,\cite{Sirunyan:2018aui} results
show a local excess of $\sim 3\, \sigma$ at
$\sim 96\,$GeV, and a similar excess of $2\, \sigma$
at roughly the same mass~\cite{CMS:2015ocq} in Run~I.
Assuming dominant gluon fusion production the
excess corresponds to
$\mu_{\rm CMS}=0.6 \pm 0.2 \; .$
First Run\,II~results from~ATLAS
with~$80$\,fb$^{-1}$ in the diphoton final state turned
out to be weaker than the corresponding CMS results, see, e.g., Fig.~1
in~\cite{Heinemeyer:2018wzl}.
Possibilities are discussed in the literature of how to
simultaneously explain both excesses by a common origin.
In particular supersymmetric realizations can be found in~\cite{Biekotter:2017xmf,Domingo:2018uim,Hollik:2018yek,Biekotter:2019gtq}.
For a review we refer to Refs.~\cite{Heinemeyer:2018jcd,Heinemeyer:2018wzl}.

\section{The N2HDM}
We discussed in~\cite{Biekotter:2019kde,Biekotter:2019mib}
how a $\sim 96\,$GeV Higgs boson of the Next to minimal 2 Higgs
Doublet Model (N2HDM)~\cite{Chen:2013jvg,Muhlleitner:2016mzt} can be the
origin of both excesses in the type II and type IV scenarios.
The N2HDM extends the CP-conserving 2 Higgs Doublet Model (2HDM) by
a real scalar singlet field. In analogy to the 2HDM, a $Z_2$ symmetry
is imposed to avoid flavor changing neutral currents at the tree level,
which is only softly broken in the Higgs potential. Furthermore, a
second $Z_2$ symmetry, under which the singlet field changes the sign,
constraints the scalar potential. This symmetry is broken spontaneously
during electroweak symmetry breaking (EWSB), as soon as the singlet
field obtains a vacuum expectation value (vev).

In total, the Higgs sector of the N2HDM consists of 3 CP-even
Higgs bosons $h_i$, 1 CP-odd Higgs boson $A$, and 2 charged
Higgs bosons $H^\pm$. In principle, each of the particles
$h_i$ can account for the SM Higgs boson at $125\,$GeV.
In our analysis, $h_2$ will be identified with the SM
Higgs boson, while $h_1$ plays the role of the potential
state at $\sim 96\,$GeV. The third CP-even and the CP-odd
states $h_3$ and $A$ were assumed to be heavier than
$400\,$GeV to avoid LHC constraints. The charged Higgs-boson
mass was set to be larger than $650\,$GeV to satisfy 
constraints from flavor physics observables.

In the physical basis the 12 indepentend parameters
of the model are
the mixing angles in the CP-even
sector $\alpha_{1,2,3}$,
the ratio of the vevs
of the Higgs doublets $\tan\beta=v_2/v_1$,
the SM vev $v=\sqrt{v_1^2+v_2^2}$,
the vev of the singlet field $v_S$,
the masses of the physical Higgs bosons
$m_{h_{1,2,3}}$, $m_A$ and $M_{H^\pm}$, and
the soft $Z_2$ breaking parameter $m_{12}^2$.
Using the public code \texttt{ScannerS}~\cite{Coimbra:2013qq,Muhlleitner:2016mzt}
we performed a scan over the following
parameter ranges:
\begin{align}
95 \gev \leq m_{h_1} \leq 98 \gev \; ,
\quad m_{h_2} = 125.09 \gev \; ,
\quad 400 \gev \leq m_{h_3} \leq 1000 \gev \; , \notag \\
400 \gev \leq m_A \leq 1000 \gev \; ,
\quad 650 \gev \leq \MHp \leq 1000 \gev \; , \notag\\
0.5 \leq \tan\beta \leq 4 \; ,
\quad 0 \leq m_{12}^2 \leq 10^6 \gev^2 \; ,
\quad 100 \gev \leq v_S \leq 1500 \gev \; . \label{eq:ranges}
\end{align}

The following experimental and theoretical constraints were
taken into account:
\begin{itemize}[noitemsep,topsep=0pt]
	\item[-] tree-level perturbativity and global minimum 
				conditions
	\item[-] Cross-section limits from collider searches
	using \texttt{HiggsBounds v.5.3.2}~\cite{Bechtle:2008jh,Bechtle:2011sb,Bechtle:2013wla,Bechtle:2015pma}
	\item[-] Signal-strength measurements of the SM Higgs boson
	using \texttt{HiggsSignals v.2.2.3}~\cite{Bechtle:2013xfa,Stal:2013hwa,Bechtle:2014ewa}
	\item[-] Various flavor physics observables, in particular 
excluding $\MHp < 650\gev$ for all values of 
$\tan\beta$ in the type~II and~IV.
	\item[-] Electroweak precision observables in terms
	of the oblique parameters $S$, $T$ and $U$~\cite{Peskin:1990zt,Peskin:1991sw}
\end{itemize}
For more details we refer to Ref.~\cite{Biekotter:2019kde}.
The relevant input for \texttt{HiggsBounds} and \texttt{HiggsSignals},
(decay withs, cross sections), were obtained using
the public codes
\texttt{N2HDECAY}~\cite{Muhlleitner:2016mzt,Djouadi:1997yw}
and
\texttt{SusHi}~\cite{Harlander:2012pb,Harlander:2016hcx}.

\section{Results}
The results of our parameter scans in the type II and
type IV N2HDM, as given in~\cite{Biekotter:2019kde}, 
%
show that both types of the N2HDM can accommodate
the excesses simultaneously, while being in agreement
with all considered constraints described above.
A preference of larger values of $\mu_{\rm CMS}$
in the type II scenario is visible, which is caused
by the suppression of decays into $\tau$-pairs
(see~\cite{Biekotter:2019kde} for details).

The particle $h_1$ is dominantly singlet-like, and
acquires its coupling to the SM particles via
the mixing with the SM Higgs boson $h_2$.
Thus, the presented scenario will be experimentally
accessible in two complementary ways. Firstly,
the new particle $h_1$ can be produced directly
in collider experiments. Secondly, deviations of
the couplings of $125\,$GeV Higgs boson $h_2$
from the SM predictions are present.
We propose
experimental analyses to constrain (or confirm)
our explanation of the excesses, both making use
of the two effects mentioned above.

\begin{figure}
  \centering
  \includegraphics[width=0.48\textwidth]{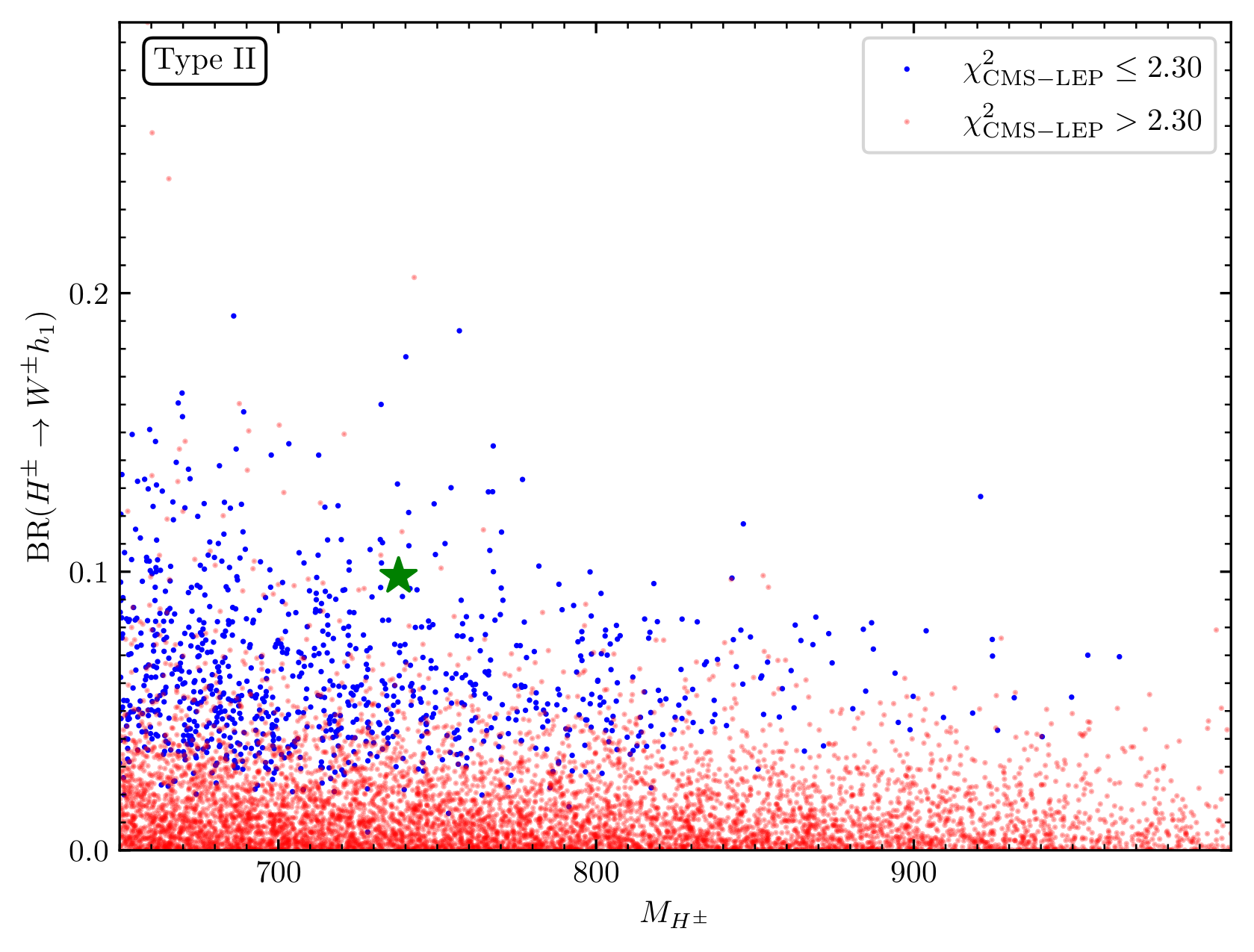}
  ~
  \includegraphics[width=0.48\textwidth]{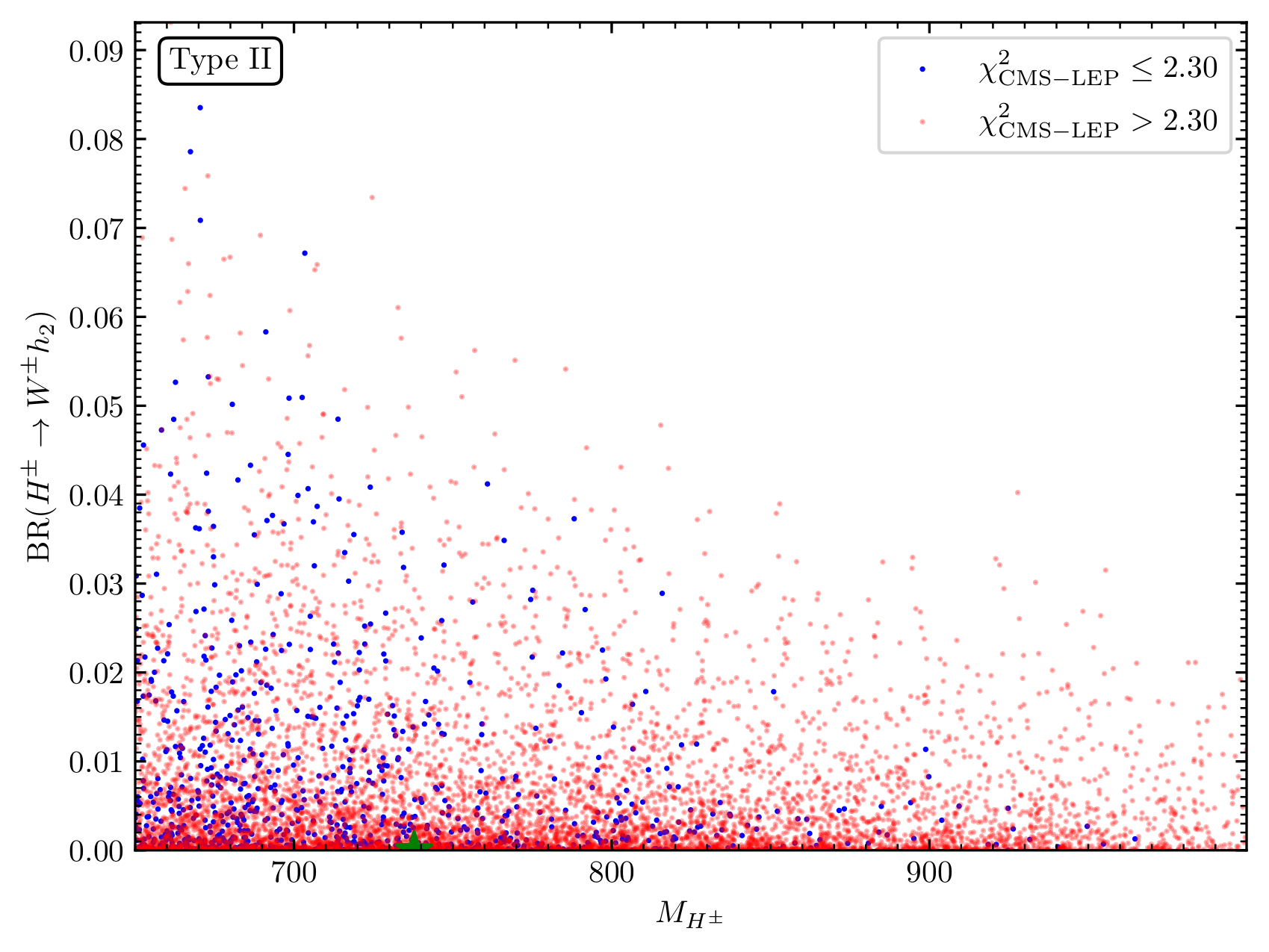}
  \caption{$\br(H^\pm \to W^\pm h_{1})$ (left) and
  $\br(H^\pm \to W^\pm h_{2})$ (right) in terms of the
  charged Higgs-boson mass $M_{H^\pm}$ for the points
  inside (blue) and outside (red) of the $1\,\sigma$ ellipse
  regarding the CMS and the LEP excesses. The best-fit point is
  highlighted with a green star.}
  \label{fig:brhpm}
\end{figure}

Due to the presence of the additional light Higgs boson
which is substantially mixed with the SM Higgs boson, 
the scenario deviates from the well-known alignment
limit of the 2HDM. One consequence is the occurrence of
sizable BRs of the charged Higgs boson
$H^\pm$ into a vector boson $W^\pm$ and a CP-even
Higgs boson, either $h_1$ or $h_2$. Kinematically, these
decays will always be allowed, because of the lower
limit of $650\,$GeV on $M_{H^\pm}$.
To demonstrate this, we show in Fig.~\ref{fig:brhpm}
the branching ratios $\br(H^\pm \to W^\pm h_{1,2})$
in the type II scenario. One can see in the left plot
that especially the points that accommodate the excesses
(shown in blue) are correlated with larger branching ratios
for the decay into $h_1$. We therefore encourage the analysis
of decays of a charged Higgs boson into a vector boson and
a light Higgs boson also outside the mass window around
$125\,$GeV for the Higgs boson in final state, as an effective
tool to search for additional light Higgs bosons.

\begin{figure}
\centering
\begin{minipage}{.44\textwidth}
  \centering
  \includegraphics[width=0.98\textwidth]{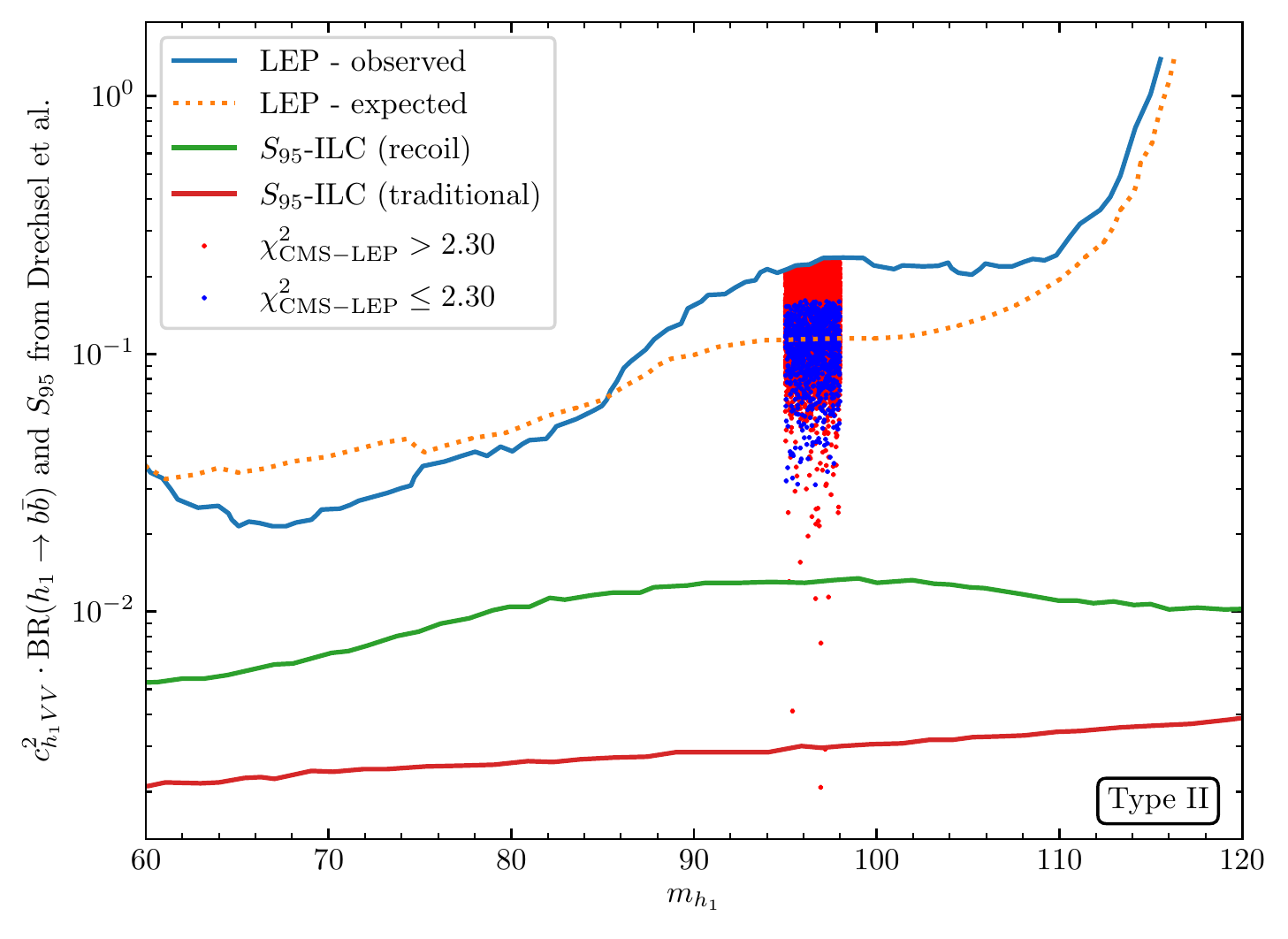}
  \caption{The $95\%$ CL expected (orange dashed) and
  observed (blue) upper bounds on the Higgsstrahlung production
  process with associated decay of the scalar to a pair of
  bottom quarks at LEP~\protect\cite{Barate:2003sz}.}
  \label{fig:2ilc}
\end{minipage}~
\begin{minipage}{.44\textwidth}
  \centering
  \includegraphics[width=0.98\textwidth]{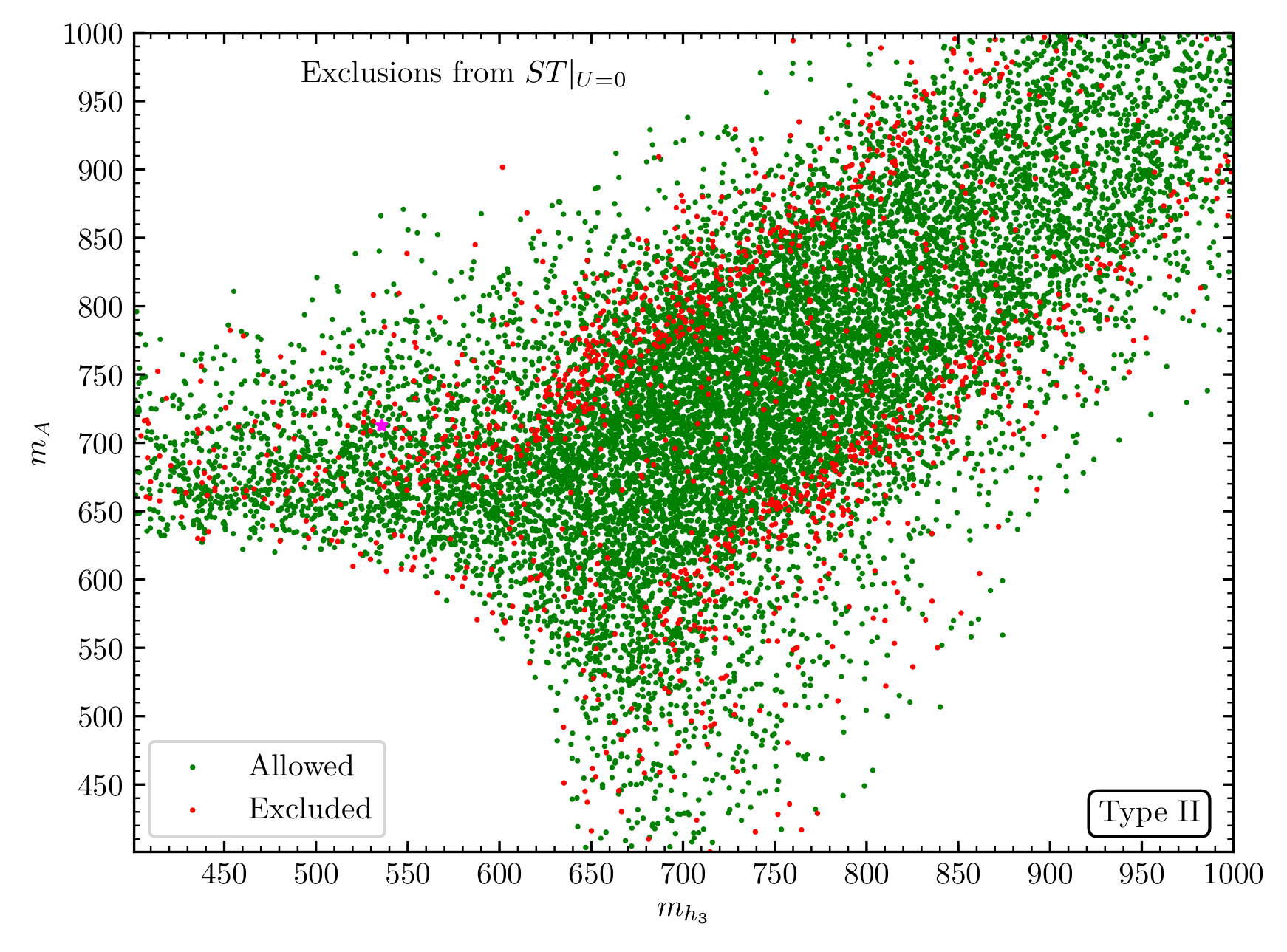}
  \caption{Points that
  fulfill the constraints on the $STU$ parameters
  (\textit{green}) in the $m_{h_3}-m_A$ plane.
  The points in \textit{red} do not fulfill
  the constraints on $ST$ with $U$ assumed to be vanishing,
  which we imposed additionally in our scans.}
  \label{fig:2stu}
\end{minipage}
\end{figure}

Regarding future collider experiments beyond the LHC, a
lepton collider is expected to be able to produce and
analyse the additional light Higgs boson $h_1$.
As an example, we compare
the current LEP bounds and the prospects of the
International Linear Collider (ILC), based on~\cite{Drechsel:2018mgd},
to our scan points 
in the type II scenario in Fig.~\ref{fig:2ilc}\,(left).
We show the expected $95\%$ CL upper limits at
the ILC using the traditional (red) and the
recoil technique (green)~\cite{Drechsel:2018mgd}. We indicate the points
  which lie within (blue) and
  outside (red) the $1 \, \sigma$ ellipse
  regarding $\mu_{\rm LEP}$ and $\mu_{\rm CMS}$.
Remarkably, all the points we found
that fit the LEP and the CMS excesses at the $1\,\sigma$
level would be excluded by the ILC, if no deviations from
the SM background would be observed. On the other hand,
if the $96\,$GeV Higgs boson is realized in nature, the
ILC would be able to produce it in large numbers.

Finally, more precise measurements of the oblique
parameters $S$ and $T$ would be an effective tool
to further constrain the neutral Higgs-boson masses
in relation to $\MHp$, see Fig.~\ref{fig:2stu}(right).
One can observe the well known correlation between charged and CP-odd
Higgs mass. However, no preference w.r.t.\ the electroweak precision data
in comparison to the two excesses can be seen; ``allowed'' and ``excluded'' points overlay each other.


\section*{References}
\bibliography{blois}







\end{document}